\documentclass[aps,prl,twocolumn,superscriptaddress,showpacs]{revtex4}

\usepackage{graphicx}% Include figure files
\usepackage{dcolumn}% Align table columns on decimal point
\usepackage{bm}% bold math
\begin{document}
\def \lafepo{LaFePO}
\def \Ba122{BaFe$_2$As$_2$}
\def \CoBa122{Ba(Fe$_{1-x}$Co$_x$)$_2$As$_2$}
\def \YBCO{YBa$_2$Cu$_3$O$_{7-\delta}$}
\def \KBa122{Ba$_{1-x}$K$_x$Fe$_2$As$_2$}
\def \LSCO{La$_{2-x}$Sr$_x$CuO$_4$}
\def \Tc{$T_c$}
\def \Torth{$T_{ortho}$}
\def \TN{$T_N$}
\def \TS{$T_S$}

\title{Evidence for an electron nematic phase transition in underdoped iron pnictide superconductors}



%\thanks{}
\author{Jiun-Haw Chu}
\affiliation{Department of Applied Physics and Geballe Laboratory for Advanced Materials, Stanford University, Stanford, California 94305, USA}
\affiliation{Stanford Institute of Energy and Materials Science, SLAC National Accelerator Laboratory, 2575 Sand Hill Road, Menlo Park 94025,California 94305, USA}
\author{James G. Analytis}
\affiliation{Department of Applied Physics and Geballe Laboratory for Advanced Materials, Stanford University, Stanford, California 94305, USA}
\affiliation{Stanford Institute of Energy and Materials Science, SLAC National Accelerator Laboratory, 2575 Sand Hill Road, Menlo Park 94025,California 94305, USA}
\author{Kristiaan De Greve}
\affiliation{E. L. Ginzton Laboratory, Stanford University, Stanford, California 94305, USA}
\author{Peter L. McMahon}
\affiliation{E. L. Ginzton Laboratory, Stanford University, Stanford, California 94305, USA}
\author{Zahirul Islam}
\affiliation{The Advanced Photon Source, Argonne National Laboratory, Argonne, Illinois 60439, USA}
\author{Yoshihisa Yamamoto}
\affiliation{E. L. Ginzton Laboratory, Stanford University, Stanford, California 94305, USA}
\affiliation{National Institute of Informatics, Hitotsubashi 2-1-2, Chiyoda-ku, Tokyo 101-8403, Japan}
\author{Ian R. Fisher}
\affiliation{Department of Applied Physics and Geballe Laboratory for Advanced Materials, Stanford University, Stanford, California 94305, USA}
\affiliation{Stanford Institute of Energy and Materials Science, SLAC National Accelerator Laboratory, 2575 Sand Hill Road, Menlo Park 94025,California 94305, USA}


\date{\today}

\begin{abstract}
Electrical resistivity measurements of detwinned single crystals of the representative iron arsenide \CoBa122 reveal a dramatic in-plane anisotropy associated with the tetragonal-to-orthorhombic structural transition that precedes the onset of long-range antiferromagnetic order. These results indicate that the structural transition in this family of compounds is fundamentally electronic in origin.
\end{abstract}

%\keywords{}
\maketitle

It is often difficult to disentangle cause and effect in complex interacting systems.  The relationship of lattice and electronic transitions in the recently discovered family of superconducting iron arsenides provide an interesting case in point. The parent compounds are non-superconducting antiferromagnets\cite{Kamihara,Rotter} with a Fermi surface consisting of several small pockets resulting from reconstruction due to the broken translational symmetry\cite{Sebastian,Analytis}. Suppression of the antiferromagnetic ground state by various means eventually leads to superconductivity, with critical temperatures of up to 55 K\cite{Ren}. Significantly, the antiferromagnetic transition is always preceded by or coincident with a tetragonal to orthorhombic structural distortion\cite{LaFeAsO,Lester,Pratt}. It has been proposed that this structural distortion is driven by an electron nematic state\cite{Fang,Yildirim,Xu,Johannes}, which could for example arise from the fluctuating antiferromagnetism\cite{Fang,Yildirim} or orbital ordering\cite{Kruger,Lv,Chen}. Here we report measurements of in-plane resistivity anisotropy of detwinned crystals of the representative iron pnictide \CoBa122 . We find that the resistivity along the shorter $b$-axis $\rho_b$ has a negative temperature gradient, showing insulating-like behavior, while $\rho_a$ is metallic. The ratio of in-plane resistivities reaches a maximum value of two at the beginning of the superconducting dome in the phase diagram, and the fluctuations of this anisotropy extends well above the structural transition temperature. These observations provide compelling evidence that the orthorhombic structural transition is fundamentally electronic in origin, with consequences for the superconducting pairing interaction.

	Electron nematic order refers to a spontaneously broken rotational symmetry in a system of interacting electrons without the breaking of translational symmetry\cite{Fradkin}. Perhaps the most studied manifestation is in quantum Hall systems\cite{Lilly}, but there is growing evidence for electron nematic phases in Sr$_3$Ru$_2$O$_7$\cite{Borzi}, and in underdoped cuprates\cite{Ando, Hinkov}. In particular, the recent observation of a large in-plane anisotropy in the Nernst effect in \YBCO suggests that the much-debated pseudo-gap phase is in fact nematic in origin\cite{Daou}. The crucial experiments in most of these cases probe the in-plane transport anisotropy that arises due to the nematic order. In this paper we show that underdoped crystals of the representative iron arsenide \CoBa122 exhibit a dramatic in-plane resistivity anisotropy associated with the orthorhombic structural transition, indicating that electron nematic order plays an important role in this family of compounds too.

\begin{figure}
\includegraphics[width=8.5cm]{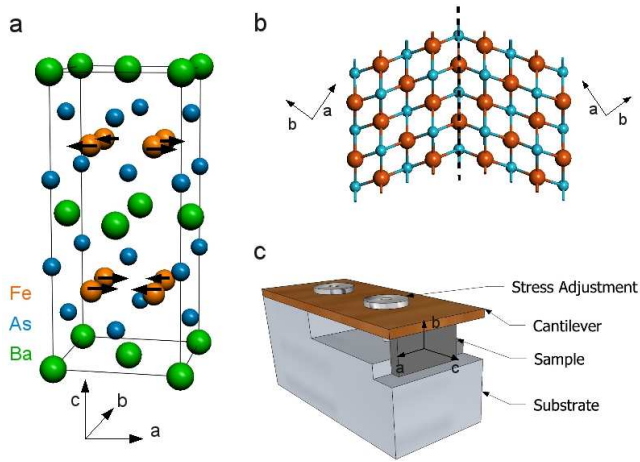}
\caption{\label{fig1}(a) Schematic diagram of the crystal structure of \Ba122 in the antiferromagnetic state. The magnetic moments on the iron sites point in the a-direction, and align anti-parallel along the longer a-axis, and parallel along the shorter b-axis. (b) Schematic diagram illustrating a twin boundary between two domains that form on cooling through the structural transition at Ts. Dense twinning in macroscopic crystals obscures any in-plane electronic anisotropy in bulk measurements. (c)  Schematic diagram of the device used to detwin single crystals in situ. The sample is held sandwiched between a cantilever and a substrate, with a screw in the center of cantilever to adjust the uni-axial pressure. The (0 0 1) surface of the crystal is exposed, enabling transport measurements.}
\end{figure}

 One difficulty with probing the in-plane electronic anisotropy of the iron arsenides is that the material naturally forms dense structural twins below the orthorhombic transition (Figure \ref{fig1}a,b)\cite{Tanatar}. Measurements of twinned samples present only an average of the intrinsic anisotropy, from which little detailed information can be extracted. For the present study we have developed a mechanical cantilever device (shown in Fig. \ref{fig1} (c)) that is able to detwin crystals in situ. Crystals were cut such that the orthorhombic $a$ and $b$ axes were aligned parallel to the direction of applied stress. With only modest pressures (approximately 5 MPa, estimated from the deflection of the cantilever), small enough that the critical temperatures associated with the structural and Neel order (\TS\ and \TN) are unaffected, the device is able to almost fully detwin underdoped crystals, revealing for the first time the previously hidden in-plane electronic anisotropy. 

\begin{figure}
\includegraphics[width=8.5cm]{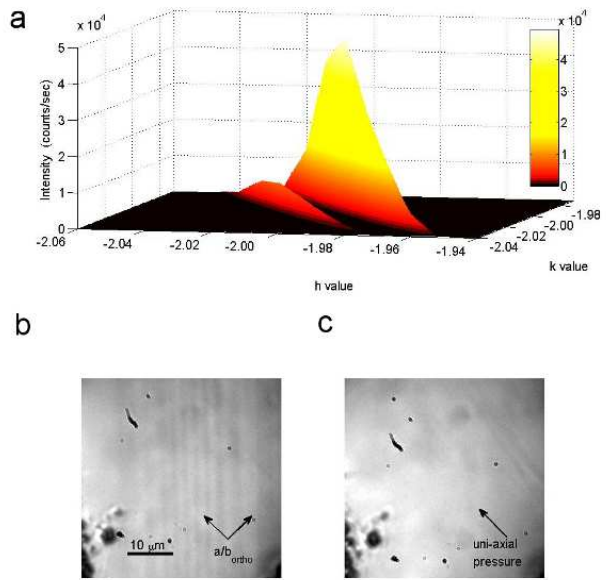}
\caption{\label{fig2}(a) Splitting of the (-2 -2 20) Bragg peak (referenced to the tetragonal lattice) at 40 K for a sample with x = 0.025 under uni-axial pressure, as revealed by high-resolution x-ray diffraction. The two dimensional surface is formed by interpolating between Gaussian fits to a dense mesh of points from h-k scans. Each peak corresponds to one of the two twin domains. The relative volume fraction of the larger domain, corresponding to the shorter $b$-axis aligned parallel to the applied stress, is approximately 86\% . (b) Polarized-light image of the surface of an unstressed \Ba122 crystal at 5K without applying uni-axial pressure. The light and dark stripe patterns revealed the existence of twin domains below structural transition. (c) Polarized-light image on the surface of the same crystal under uni-axial pressure at 5K. The absence of stripes indicates the applied pressure has detwinned the crystal.}
\end{figure}

Evidence of the efficacy of this method for detwinning can be obtained from high resolution x-ray diffraction measurements. Data shown in Figure \ref{fig2}(a) reveal the splitting of the (-2 -2 20) Bragg peak (with respect to the tetragonal lattice) in the orthorhombic state of a crystal with composition x = 0.025 held under uni-axial pressure with the cantilever. Comparison of the integrated intensity of the two peaks yields a relative volume fraction of 86\% of the twin orientation with the shorter $b$-axis along the applied stress. Direct optical images were also taken to confirm this result. Samples were illuminated by polarized light, and the reflected light was collected through an almost crossed polarizer so as to maximize the contrast due to the different birefringence of the two twin orientations. Representative images of a \Ba122 sample surface at 5K are shown for the relaxed cantilever (Figure \ref{fig2}(b)) and strained cantilever (Figure \ref{fig2}(c)). Stripes associated with the twin domains are clearly visible for the unstrained crystal, but these have completely disappeared for the strained sample.

\begin{figure*}
\includegraphics[width=17cm]{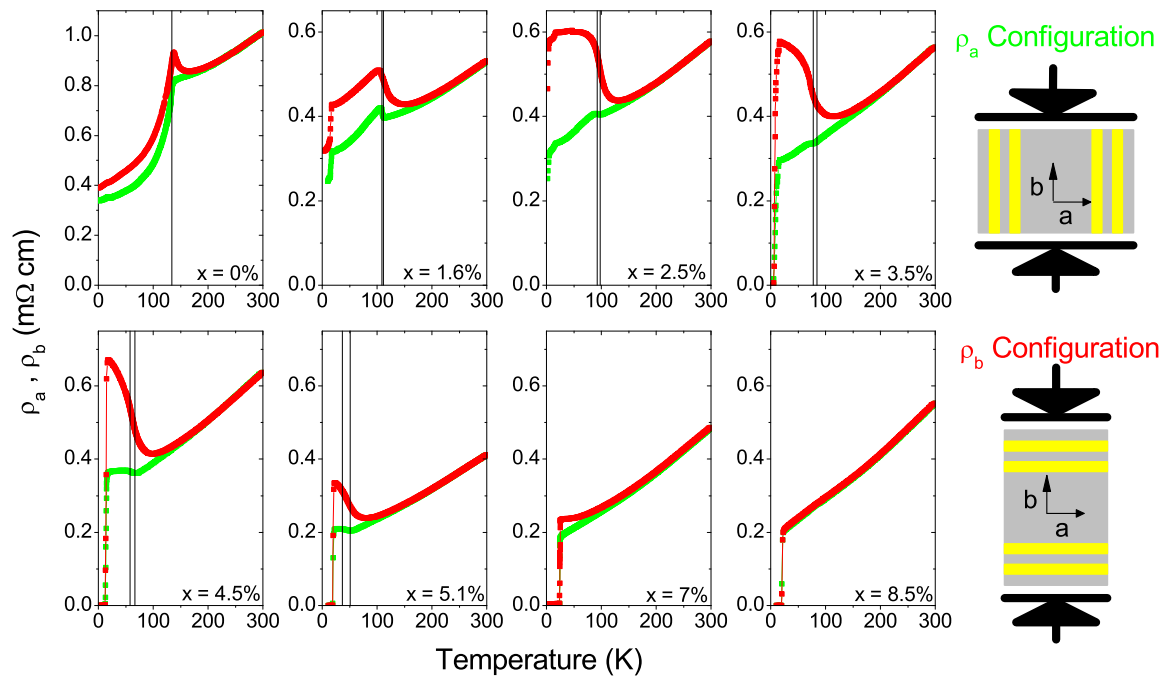}
\caption{\label{fig3} Temperature dependence of the in-plane resistivity $\rho_a$ (green) and $\rho_b$ (red) of \CoBa122 for Co concentrations from x = 0 to 0.085. Solid and dashed vertical lines mark critical temperatures for the structural and magnetic phase transitions \TS and \TN  respectively. Values are identical to those found for unstressed samples, indicating that the uni-axial pressure serves as a weak symmetry breaking field to orient twin domains without affecting the bulk properties. The uni-axial stress does, however, appear to affect the superconducting transitions in some underdoped samples, inducing a partial superconducting transition for x=0.016 and 0.025, which are not observed for unstressed crystals. Schematic diagrams in the right-hand panel illustrate how measurements of $\rho_a$ and $\rho_b$ were made. Dark arrows indicate the direction in which uni-axial pressure was applied, and smaller arrows indicate the orientation of the a and b crystal axes. In all cases, the same samples and the same contacts (shown in gold for a standard 4-point configuration) were used for both measurements. }
\end{figure*}

	Resistivity data were collected for detwinned samples as a function of temperature. Eight representative cobalt concentrations were measured, from the undoped parent compound x = 0 through to fully overdoped composition (x = 0.085). Representative data for each of these compositions are shown in Figure \ref{fig3}. Measurements were made for currents applied parallel and perpendicular to the applied stress, yielding $\rho_b$ (shown in red) and $\rho_a$ (shown in green) respectively.

\begin{figure*}
 \includegraphics[width=17cm]{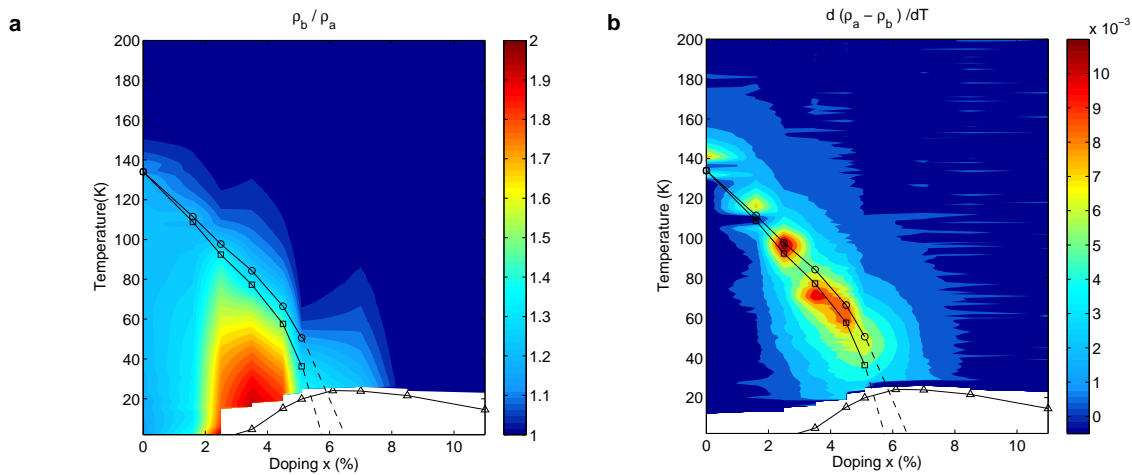}
\caption{\label{fig4}(a) Evolution of the in-plane resistivity anisotropy as a function of temperature and doping, expressed in terms of the resistivity ratio $\rho_b$/$\rho_a$. Structural, magnetic and superconducting critical temperatures, determined following Ref. \cite{CoBa122} are shown as circles, squares and triangles respectively. Significantly, the resistivity ratio deviates from unity at a considerably higher temperature than \TS , indicating that nematic fluctuations extend far above the phase boundary. (b) The difference in the temperature derivative of $\rho_a$ and $\rho_b$ ($d\rho_a/dT- d\rho_b/dT$) as a function of temperature and doping. The resistivity has been normalized by its room temperature value to avoid uncertainty due to geometric factors. Regions of highest intensity mark the regions where $\rho_b$ appears to be insulating-like ($d\rho_b/dT < 0$) while $\rho_a$ remains metallic ($d\rho_a/dT > 0$). This behavior is clearly correlated with the electronic nematic phase between the structural and magnetic transitions.}
\end{figure*}

	Inspection of Figure \ref{fig3} reveals that $\rho_b > \rho_a$ for all underdoped compositions. This result, anticipated from earlier magnetoresistance measurements\cite{AMRO}, is somewhat counterintuitive given that the collinear spin arrangement below \TN\ comprises rows of spins that are arranged ferromagnetically along the $b$-axis, and antiferromagnetically along the $a$-axis. All else being equal, one might anticipate the opposite anisotropy, and this observation hints at a more complex situation. The degree of in-plane anisotropy can be characterized by the ratio $\rho_b$/$\rho_a$, which is shown as a color scale in Figure \ref{fig4}(a). The anisotropy varies with temperature and composition, but reaches values of up to two for $0.025 \leq x \leq 0.045$ at low temperature, coincidental with the beginning of the superconducting dome in the phase diagram.  In contrast, the overdoped composition x = 0.085, which remains tetragonal for all temperatures, reveals no in-plane anisotropy.

	The temperature dependence of the resistivity is especially striking. At high temperatures, the resistivity is isotropic and almost linear over the limited temperature range investigated. For currents running in the $b$-direction the resistivity deviates from this linear behavior at a temperature significantly above Ts and increases steeply with decreasing temperature. This insulating-like behavior is cut off near \TN\ for the lowest doping levels, but extends to much lower temperatures for larger cobalt concentrations. In contrast, for currents flowing in the a direction, the resistivity behaves essentially like that of a normal metal, continuing to decrease with decreasing temperature over the entire temperature range, except for a small jump near \TN . The difference of temperature derivatives of $\rho_b$ and $\rho_a$ (normalized by the room temperature value), shown as a color scale in Figure \ref{fig4}(b), illustrates how this effect evolves across the phase diagram, revealing a strong correlation with the orthorhombic distortion.

	The composition dependence of the anisotropy in the in-plane resistivity depicted in Figure \ref{fig4}(a) and (b) is in stark contrast to that of the structural anisotropy that develops below \TS . The orthorhombic distortion, characterized by the ratio of in-plane lattice constants, has a maximum value for $x$=0 at low temperature of $(a-b)/(a+b)$ = 0.36\%, and decreases monotonically with increasing cobalt concentration\cite{Prozorov}. In sharp contrast, the in-plane resistivity develops a more pronounced anisotropy as the Co concentration is increased towards optimal doping. Clearly, the itinerant electrons are profoundly affected by this phase transition, and in a manner that is much less apparent in the response of the crystal lattice, providing compelling evidence that the underlying electronic system plays the dominant role.
 
	Significantly, the effects of this phase transition are evident for temperatures well above \TS . Even though the crystal symmetry is tetragonal above \TS , the uni-axial stress applied by the cantilever breaks the 4-fold symmetry in the basal plane, revealing the presence of the incipient electronic anisotropy. It is unlikely that this effect reflects the presence of a third phase transition at a temperature above \TS , because the phase transition would have to break exactly the same rotational symmetry that is broken explicitly at \TS . In addition, despite clear evidence in both thermodynamic and transport properties for divergent behavior associated with phase transitions at \TN and \TS , there is no evidence for divergent behavior at the onset of the resistive anisotropy. Rather, it is more likely that the effect is associated with nematic fluctuations above \TS . In this case, the critical temperature must have been suppressed considerably below the mean field temperature, implying a quasi-low dimensional character to the system.

Co-doped \Ba122 was chosen for this initial study because the structural and magnetic transition are clearly separated in temperature, and the material can be readily tuned in a controlled and reproducible fashion. However, given the rather generic phase diagram found in this family of compounds\footnote{For the Fe arsenides and oxy arsenides typified by BaFe2As2 and LaFeAsO, the structural transition is from tetragonal to orthorhombic symmetry. The closely related Fe-chalcogenides, typified by FeSe, also suffer a structural distortion, but in this case from orthorhombic to monoclinic\cite{Li}}, it is likely that the observed anisotropy is quite general to underdoped Fe-pnictides. For example, recent STM measurements also reveal a significant electronic anisotropy of Co-doped CaFe$_2$As$_2$ at low temperatures\cite{Chuang}.

It is intriguing to compare broad features associated with the Fe-pnictides with those of the more complex high-Tc cuprates. Superconductivity in both families of compounds is associated with the suppression of an antiferromagnetic ground state. And in both cases, evidence has been advanced suggesting a succession of broken rotational (nematic) and translational symmetry on the underdoped side of the phase diagram that is driven by the underlying electronic system\cite{Ando,Hinkov,Daou}. Significantly, the Fe-arsenides have a well-defined structural phase transition associated with the nematic phase transition. In this sense, the Fe-pnicitides present a cleaner system in which to investigate the physical origin of electron nematic order and the consequences for superconductivity.

If the two phase transitions shown in Figure \ref{fig4} are truly second order, and remain so to $T$=0, then the superconducting dome hides two quantum critical points, one magnetic and one nematic, each of which is directly associated with the same itinerant electrons from which the superconductivity is born. The relative roles played by these remains to be established, but preliminary theoretical work based on a simplified two-orbital model has shown that orbital fluctuations can enhance pairing from spin fluctuations due to the pronounced orbital character of the Fermi surface\cite{Zhang}, implying that the nematic order is not innocent in high temperature superconductivity in this family of compounds, but might in fact play a key role. 

The authors thank C. D. Batista, C.-C. Chen, T.P. Devereaux, S.A. Kivelson, A. P. Mackenzie, R.D. McDonald, S.C. Riggs, D. J. Scalapino and Z.-X. Shen for helpful discussions. This work is supported by the DOE, Office of Basic Energy Sciences, under contract no. DE-AC02-76SF00515


%\bibliography{Ba122AMROv6}

\end{document}